\documentclass[a4paper,11pt]{article}
\usepackage{jcappub} % for details on the use of the package, please see the JINST-author-manual
\usepackage{graphicx} % Required for inserting images
\usepackage{amsmath}
\usepackage{bm}
\arxivnumber{ } % Only if you have one
\title{Microlensing of halo objects in the exterior part of the Galaxy}

\author[a]{Tabib Rayed Hossain}
\author[b]{Prabir Kumar Haldar}
\author[c]{Mehedi Kalam}

\affiliation[a]{Department of Physics,Amity University Kolkata,\\ Action Area II, Rajarhat,
Kolkata-700135, India. }
\affiliation[b]{Department of Physics, Coochbehar Panchanan Barma University,\\ Coochbehar,
West Bengal, India.}
\affiliation[c]{Department of Physics,Aliah University,\\ II-A/27, Action Area II, Newtown,
Kolkata-700160, India.}

\emailAdd{tab.rh20@gmail.com}
\emailAdd{prabirkrhaldar@gmail.com}
\emailAdd{kalam@associates.iucaa.in(corresponding author)}

\abstract{In the context of this paper, microlenses present as oblate clusters of dark matter structures called massive astrophysical compact halo object(MACHO) in the
galactic halo are considered. The NFW density profile \cite{navarro1997universal} is derived from the observational data and works
best in the halo region of the exterior part of the galaxy. Hence this profile is used to plot the potential, deflection angle, and
critical and caustic curves for the aforementioned microlenses using numerical methods. Moreover, this model is compared with
an older density profile model \cite{bogdanov2001}, and the differences in their caustic and critical curves are pointed out. This leads
to the conclusion that the NFW model produces caustic and critical curves that occupy a smaller region than portrayed by the
caustic and critical curves produced by the older density profile model. However, the differences are not that significant to these
structures act as microlenses as they are so small and beyond the scope of modern telescopes, and thus only the light curves can
be detected from these structures.}

\begin{document}
\maketitle
\flushbottom

\section{Introduction}
The proposition of the existence of dark matter came about when observations suggested that a lot of matter (above 90 percent) is made up of invisible matter which means it does not interact or give off radiation which can be detected using a telescope. So it is not known what particles are they made of. This matter is thus called dark matter as it is dark or invisible to the telescopes available. Brief history and the advent of this topic to the fore front can be found in the papers by Bertone \cite{bertone2018history} and Garrett\cite{garrett2011dark}. Analyzing nucleosynthesis and galaxy formation leads to the conclusion that these particles are non-baryonic and interact weakly with themselves as well as baryonic particles. But an interesting phenomenon they exhibit is that they have mass and thus have gravity and thus bend the space-time around them. Hence it interacts with everything around it, through its gravitation. These are also called non-dissipative dark matter. These dark matter are proposed to be made up of WIMPS which are a hypothetical particles. These particles are being tried to be detected which would provide a direct proof of the existence of dark matter as in the case \cite{schumann2019direct} and \cite{cushman2013working}.

This type of dark matter plays a crucial role in the formation of the massive structures present in the universe like galaxies, galaxy clusters, and superclusters. As suggested by Gurevich \cite{gurevich1997} during the early stage of development, small perturbations experience gravitational compression and subsequent mixing of dark matter, resulting in the creation of stationary self-trapped objects. This can also be corroborated by Berezinsky \cite{berezinsky2014small}. The normal matter which is a very small part of the whole mass of the matter is present in the center where the density is at its peak and the dark matter is located according to the law (1). This theory was confirmed by the data \cite{burstein1985,gurevich1995,persic1995}.

\begin{equation}
    \rho(r)= kr^{-\alpha}
    \label{eq1}
\end{equation}

Here, $\alpha=1.8$. Gravitational attraction brings several galaxies to come together to form clusters which can in turn form larger clusters called Abell clusters and super-clusters . This leads to a hierarchical clumping of massive structures and an important role is played by dark matter for this process. This idea can be extended to the small scale as previously it was assumed that the smallest objects that can appear were in the range of $(10^7 - 10^8)M_{\odot}$ in halos of small galaxies. But Gurevich \cite{gurevich1995a,dremin1997second} showed tiny structures can be created by the dark matter which is bound by gravity.

However, the question is if these small-scale objects are present now or were present in the early universe and if they are at all present in what form. But a detailed study \cite{dremin1997second} determines that the lifespan of these small-scale entities is greater than the age of the universe. An important study \cite{alcock1995} suggests the existence of a vast number of small-scale invisible objects of mass $M\approx(0.05-0.8)M\odot$ in our galactic halo. They were previously hypothesized to be brown or white dwarfs or Jupiter-like planets made of normal matter which are just very dim. But another study \cite{alcock1997macho} that the observed entities constitute approximately 50 percent of the overall dark matter content within our galactic halo. This leads to the belief that some of these small-scale objects are made of dark matter if not all. A recent study \cite{brito2022snowmass2021} suggest that dark matter small-scale objects may be revolutionary in studying the particle physics aspect of it. There has been increase in the number of studies studying dark matter in the small scale in pulsars as in the following studies: \cite{molla2020does}, \cite{rahman2020possible} and \cite{rahman2022possibleexistence}. Though the shapes of the flux variation curves can be explained by a simple Schwarzschild with a symmetric profile \cite{gurevich1995} in some cases it can be better explained by a non-compact lens model \cite{sazhin1996microlensing}. 

There were various missions conducted to observe lensing events one of them were MACHO which was aimed at finding dark matter, other than that mission like OGLE, EROS, PLANET collaborations. Many of these projects were aimed at detecting exoplanets. But MACHO and OGLE were instrumental in detecting microlensing events to detect MACHOS. Most instrumental in these were OGLE-I ( before 1998 ), OGLE-II (1998-2002), OGLE-III (2002-2009), OGLE-IV (2011-2022) but OGLE-IV was continued in 2022 and is still ongoing. Calchi\cite{calchi2011microlensing} studied results OGLE-II and OGLE-III which were observing lenses in our Galactic Halo using stars in the Large Magellanic Cloud. He concluded that if we take the MACHO mass in the range $10^{-2} - 0.5M_{\odot}$, then the MACHOs make up about $10-20$ percent of all the dark matter in theHalo of the Milky Way. This Percentage also called the halo mass fraction($f$) and for masses in the order of $1.0M_{\odot}$ it is around $24$ percent. The upper limit to $f$ is around $50$ percent which is the largest ever found for a mass of $100M_{\odot}$.

The mass range in reference to this paper shows a picture such that these structures are rare to find and thus very hard but this cannot be ruled out. Using a good and accurate model can be instrumental in making detection easier. Thus in this paper we will attempt to do so. The goal of this paper is to fit the Navarro–Frenk–White(NFW) dark matter density profile to the non-singular power law and find the Caustic and critical curve for the above case. The NFW dark matter density profile is a newer model of Dark matter suggested by Julio Navarro, Carlos Frenk, and Simon White \cite{navarro1997universal} and works really well in the halo region of our galaxy. The earlier model suggested previously by Gurevich\cite{gurevich1997} is a theoretical model which was fitted there on the other hand NFW model is derived from the galactic rotation curve and is a better representation.

\section{NFW Dark Matter Density Profile}

The Density profile suggested by Gurevich \cite{gurevich1997} was somewhat proposed in 1997 and is an old model for dark matter so in this paper, we would use the (NFW) Dark Matter Density Profile \cite{navarro1997universal}. The NFW Density model can be written as the following:
\begin{equation}
    \rho (r)={\frac {\rho _{0}}{{\frac {r}{r_{0}}}\left(1~+~{\frac {r}{r_{0}}}\right)^{2}}}
    \label{eq2}
\end{equation}
In the above equation, $r_{0}$ is the core radius and $\rho_{0}$ is the effective core density.  This model is valid in the outer regions of the halo. But we will miniaturize this model for a small-scale system in the range of $M = (0.05-0.8)M_{\odot}$. 

To calculate the equation of the core mass $M_{x}$ a volumetric integration is needed to be calculated. This equation can be used to calculate the effective core density $\rho_{0}$.

\begin{equation}
    M_{x}=\int _{0}^{{R_{x}}}4\pi r^{2}\rho (r)\,dr=4\pi \rho _{0}R_{0}^{3}\left[\ln \frac{R_{0}+R_{x}}{R_{0}} - \frac{R_{x}}{R_{x}+R_{0}} \right]
    \label{eq3}
\end{equation}

The above equation is used to calculate all the graphs in this paper. It can be noted that the spherical symmetry of WIMPS may break down due to mainly two reasons. One of them is that they may get deformed due to the gravitation influence of other objects like other clusters or stars of the same galaxy as pointed out by Gurevich \cite{gurevich1997}. But in this case, as we consider a small-scale model we can ignore that effect as its probability is quite small. The other reason the symmetry may break down is due to the rotation of the system and this seems more likely. Hence we consider an oblate system due to the rotation of such particles.

\section{Gravitational Lensing Formalism}

The effect of Gravitational Lens for the assumed object depends on the surface mass density of the object which is denoted by the $\Sigma(\xi)$ and here $\xi$ is a vector defining the position of any point on the plane of the lens. It is needless to say all the vectors below are in bold. In a coordinate system with its origin aligned with the center projection of the particle cluster onto the lens plane, considering a spherically symmetric scenario, we can define the surface mass density as:

\begin{equation}
\Sigma(\xi) = 2\int_{0}^{\sqrt{R_{x}^2-\xi^2}} \rho(\sqrt{\xi^2+z^2}) dz
\label{eq4}
\end{equation}

The above surface mass density has rotational symmetry. But for simplicity, we define a dimensionless surface mass density in the following way:

\begin{equation}
    \kappa(\xi) = \frac{\Sigma(\xi)}{\Sigma_{cr}}
    \label{eq5}
\end{equation}

where, $\Sigma_{cr}$ is called the critical density. Critical density can be written as :

\begin{equation}
    \Sigma_{cr} = \frac{c^2 D_{S}}{4\pi GD_{d}D_{ds}}
    \label{eq6}
\end{equation}

In the above relation, $G$ is the universal gravitation constant, $c$ is the speed of light, $D_{s}$ is the observer's distance from the source, $D_{d}$ is the observer's distance from the lens and $D_{ds}$ is the source's distance from the lens.

To make things more simple we can use the cluster radius and use it to scale and introduce dimensionless vectors in the lens plane $\textbf{x}=\xi / \xi_{0}$ and in the source plane  $\textbf{y}= \eta / \eta_{0}$ where $\eta_{0}= \xi_{0} D_{s}/ D_{d}$. Also, note that in the above-stated relations $\xi_{0}=R_{x}$ which is the cluster radius. Also, we can consider the following:
\begin{equation}
    \textbf{x}=\textbf{x}(x_{1}, x_{2})
    \label{eq7}
\end{equation}

\begin{equation}
    \textbf{y}=\textbf{y}(y_{1}, y_{2})
    \label{eq8}
\end{equation}

we can use a similar method to scale the deflection angle $\bm{\alpha}(\mathbf{x})$ into a normalized vector,

\begin{equation}
    \bm{\alpha}(\mathbf{x})=\frac{D_{d}D_{ds}}{\xi_{0}D_{s}}\bm{\alpha}^0(\xi_{0} \mathbf{x})
    \label{eq9}
\end{equation}

Thus, the new lens equation can be written the following way:

\begin{equation}
    \mathbf{y} = \mathbf{x} - \bm{\alpha}(\mathbf{x})
    \label{eq10}
\end{equation}

Also  the above deflection angle $\bm{\alpha}(\mathbf{x})$ can be related to the potential using the following relation:

\begin{equation}
    \bm{\alpha}(\mathbf{x})=\nabla \psi(\mathbf{x})
    \label{eq11}
\end{equation}

But in the algorithm given by Barkana\cite{barkana1998} we need the relation for the calculation of the potential from the deflection angle which can be given by :
\begin{equation}
    \psi(x)= \frac{1}{\pi} \int d^{2}x' \kappa(x') \ln|x-x'|
    \label{eq12}
\end{equation}
Moreover, the above equation is the solution to the Poisson equation given below:
\begin{equation}
    \nabla^{2}\psi(x)=2\kappa(x)
    \label{eq13}
\end{equation}

\begin{figure}
	\includegraphics[width=\columnwidth]{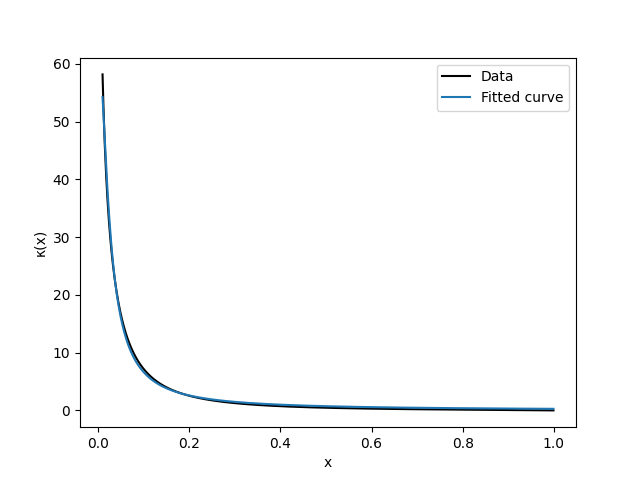}
    \caption{Distribution of dimensionless surface mass density along with the non-singular power-law elliptical model fitted onto it with parameters $q=0.28$, $u=0.020$ and $n=0.69$}
    \label{fig1}
\end{figure}

For now, these are the few formalisms required to do the following numerical analysis. Equation \eqref{eq4} can be estimated using the numerical method of integration with an accuracy of $10^{-5}$ using $R_{0}\approx0.07R_{x}$ and the central density is taken as $\rho_{0}=1.29 \times 10^{-7}g/cm^{3}$, where, $R_{x}\approx 10^{14}cm$. Also, $\rho_{0}$ was calculated using the relation in equation \eqref{eq3} where $M_{x}=0.5M_{\odot}=10^{33}g$. Whereas the $\Sigma_{cr}=5.3 \times 10^4 g/cm^{2}$ calculated for LMC by an object in our Galactic halo ($D_{d} = 8$ kpc, $D_{s} = 46$ kpc, and $D_{ds} = 38$ kpc). Using the above constants and after estimating the integration we get $\kappa(x)$ as plotted in figure \ref{fig1}. Here the dimensionless surface density at the center is scaled to one and hence the other values are scaled accordingly. 

Now, in this case, we consider an oblate shape of the whole system arising from its rotation, rather than it being spherically symmetrical. The study of the structure of rotating celestial systems assumes it to be a fluid that is both viscous and constant volume in nature which is not the case for WIMPS which we consider in this case. However, due to our assumption of a compact core and a swift decline in density with increasing distance from the center, we can employ the density distribution from the initial approximation of the Roche model. This model presumes that the complete mass of the rotating object is centered at its core. The oblateness of such objects is fairly small and the ratio of two axes is taken to be $2/3$.

Non-singular power-law elliptical distribution of surface mass density is given by: 

\begin{equation}
    \kappa(\mathbf{x}) = \frac{q}{(u^{2}+x_{1}^{2}+ex_{2}^{2})^{n}}
    \label{eq14}
\end{equation}

In the above model, $u$ is the constant core density. In the above equation, the contours of the same surface density are in the shape of an ellipse with $e=b/a$.

After estimation of equation \eqref{eq4} and calculating $\kappa(\mathbf{x})$ using the parameters as described above, $\kappa(\mathbf{x})$ can be fitted with the model given in equation \eqref{eq14} and the parameters $q$, $u$ and $n$ can be calculated by setting $e=1$. The parameters came out to be $q=0.28$, $u=0.020$ and $n=0.69$. The fitted curve is also shown in figure \ref{fig1}. Note that in the curve shown, the max value of the surface density is scaled to one and the other values are scaled accordingly.

\section{Computation of Curves}
For the purpose of simplification, it can be assumed that the cluster's axis of rotation is situated in the lens plane. Additionally, the major axes of the ellipses with equal surface density create intervals along the $x_1$ axis. Barkana \cite{barkana1998} devised an algorithm in FORTRAN called FASTELL which is available on his website and calculates the angle of deflection, Jacobi Matrix, the potential, and the amplification factor. This algorithm was ported to Python by Simon Birrer. This was used to plot the potential, deflection angle, Caustic, Critical, and magnification curves. The Caustic and critical curves can be plotted by finding the set of values of $\mathbf{x}$ and  $\mathbf{y}$ for which the amplification factor is zero and hence the magnification is infinite. The set of $\mathbf{x}$ forms the critical curve and the set of $\mathbf{y}$ values form the Caustic curve.  

\subsection{Computation of Deflection Angle}

\begin{figure}
	\includegraphics[width=\columnwidth]{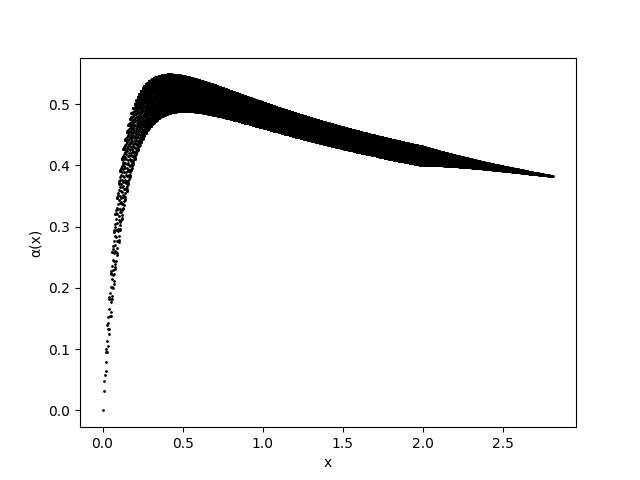}
    \caption{The deflection angle corresponding to the dimensionless mass density mentioned in the previous section. The parameters are $q=0.28$, $u=0.020$ and $n=0.69$}
    \label{fig2}
\end{figure}

The deflection angle is calculated using the FASTELL port for Python called FASTELL4py. A set of $\mathbf{x}$ values are taken in the $x$ plane and these values are iterated through the 'fastell4py.f-
astelldefl' function to calculate the deflection angle for each point. This was then plotted using the 'matplotlib' repository of Python. The parameters of equation \eqref{eq14} which are used to plot the graph are the following: $q=0.28$, $u=0.020$, and $n=0.69$. The deflection angle was plotted in fig \ref{fig2}

\subsection{Plotting Critical and Caustic Curves}

\textbf{Critical Curve}

\begin{figure}
	\includegraphics[width=\columnwidth]{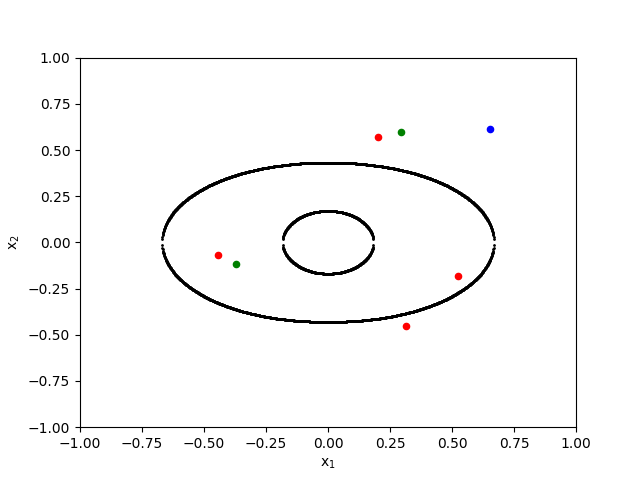}
    \caption{Critical curves or the points where the Magnification is very high. Some points are marked with different colors which can be corresponded to the Caustic Curve. The parameters are $q=0.28$, $u=0.020$ and $n=0.69$}
    \label{fig3}
\end{figure}

The critical curves are calculated by taking a set of $\mathbf{x}$ values i.e. a set of  $x_{1}$ and $x_{2}$ and calculating the determinant of the Jacobi Matrix. Next, it is checked if the determinant is zero and thus the Magnification is mathematically infinite or rather physically very high. However, the numerical method cannot produce an exact zero as the determinant of the Jacobi Matrix. But every time the value reaches zero and crosses it, it switches the sign after that point. So the written algorithm detects this switch and this point is plotted. This is iterated throughout the plane. These values are now plotted using 'matplotlib'. This plot is shown in fig \ref{fig3}

\textbf{Caustic Curve}

The Caustic Curve is calculated in a similar manner but first, the same process that is done for the plotting of the Critical Curve is repeated and those points that are detected by the algorithm are taken and the deflection angle $\alpha(x)$ for each point is calculated and then with that, $y$ is calculated as in the relation in equation \eqref{eq9}. By extending the idea of equations \eqref{eq7} and \eqref{eq8}, the following relations can be given:

\begin{equation}
    y_{1}=x_{1}-\alpha_{1}
    \label{eq15}
\end{equation}

 \begin{equation}
     y_{2}=x_{2}-\alpha_{2}
     \label{eq16}
 \end{equation}

\begin{figure}
	\includegraphics[width=\columnwidth]{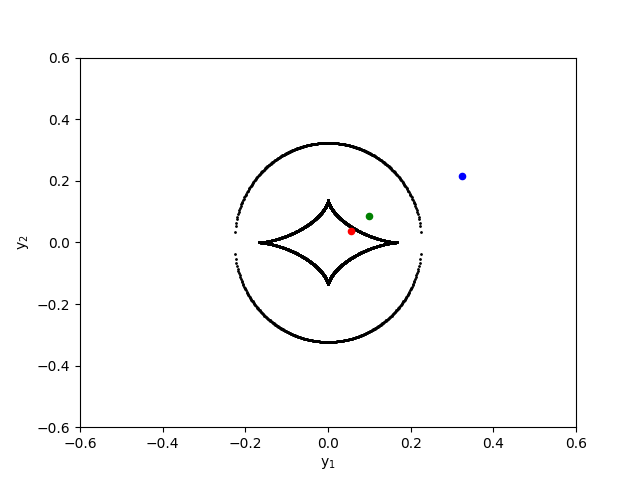}
    \caption{Caustic curves or the points where the Magnification is very high. Some points are marked with different colors which can be corresponded to the same points which are mapped from the Critical Curves. The parameters are $q=0.28$, $u=0.020$ and $n=0.69$}
    \label{fig4}
\end{figure}

Using this process the set of y values are calculated for which the magnification is infinite. This is now plotted using 'matplotlib' and this plot is shown in fig \ref{fig4}.

In a single-point lens case, it can be observed that it produces two images. But in this case, which is an extended and elliptical system, it has the ability to create multiple images in the plane of the lens. This is shown in this case in the figures \ref{fig3} and  \ref{fig4}. It can be observed that 3 points are color-coded by red ($y_{1} = 0.055, y_{2}=0.036$), green ($y_{1} = 0.100, y_{2}=0.085$) and blue ($y_{1} = 0.325, y_{2}=0215$) in the source plane (figure \ref{fig4}). But this mapped onto the lens plane (figure \ref{fig3}) using the same colors and see that some points produce more than one image in the plane where the lens is situated. The images of the source star are calculated by solving the lens equation \eqref{eq10} with an accuracy of 0.001 using Python.
\\
\\
\\
The table \ref{tab1} contains data about the marked points in fig (\ref{fig3}) and (\ref{fig4}).
Note that in the above table $\mu$ is the gravitational lens amplification factor given by the following equation:
\begin{equation}
    \mu=\frac{1}{|\det{A}|}
    \label{eq17}
\end{equation}

Upon a point in the source plane crossing the caustic, the factor of amplification or simply the magnification can mathematically reach infinity as $\det{A}$ reaches zero. But practically it can reach very high values as the source has a finite angular size. The three points in the caustic curve and their respective images in the critical curve are as shown in figure \ref{fig3} and \ref{fig4} are mentioned in the above table along with the value of their $\det{A}$ i.e. the Jacobian Matrix determinant and the magnification of the gravitational lens ($\mu$). The final column represents the type of image.

There are two types of images :
\begin{itemize}
    \item Type I: This is called positive parity or direct image i.e. when $\det{A}>0$.
    \item Type II: This is called negative parity or inverted image i.e. when $\det{A}<0$.
\end{itemize}
A physicist named Burke \cite{burke1981} proposed a theorem that suggests that "the number of images produced by a transparent gravitational lens with a limited deflection angle must always be odd".
It can be seen that the second and third sets of $y$ values produce even images that seem to contradict the theorem. Bray \cite{bray1984} showed that this happens as two images combine together and become unresolved due to the fact that there is finite accuracy while calculating the angle of deflection $\bm{\alpha}(x)$. 

In this section, the potential curve, the deflection angle curve, and the critical and caustic curves are plotted and explained how it is plotted. The different points in the plane of the source and their image in the plane of the lens are also shown in figures \ref{fig3} and \ref{fig4} and then organized in the table \ref{tab1}. Near these caustic and critical curves magnification is maximum and hence maximum deflection in the light curve.

\begin{table}
\centering
\begin{tabular}{p{2cm}|p{2cm}|p{2cm}|p{2cm}|p{2cm}|p{2cm}}
     \hline
     No.  &  $x_{1}$  &  $x_{2}$  &  detA  &  $\mu$  &  Type\\
     \hline 
     \multicolumn{6}{c}{Source 1: $y_{1} = 0.055, y_{2}=0.036$ }\\
     1.  &  $+0.652$   & $+0.611$   &  $+0.480$ &  $2.081$  & T-I\\
     \multicolumn{6}{c}{Source 2: $y_{1} = 0.100, y_{2}=0.085$ }\\
     1.  &  $-0.369$   & $-0.115$   &  $-0.590$ &  $1.694$   & T-II\\
     2.  &  $+0.296$   & $+0.595$   &  $+0.340$ &  $2.939$   & T-I\\
     \multicolumn{6}{c}{Source 3: $y_{1} = 0.055, y_{2}=0.036$ }\\
     1.  &  $-0.442$   & $-0.068$   &  $-0.488$ &  $2.048$  & T-II\\
     2.  &  $+0.200$   & $+0.572$   &  $+0.284$ &  $3.509$  & T-I\\
     3.  &  $+0.313$   & $-0.451$   &  $+0.129$ &  $7.697$  & T-I\\
     4.  &  $+0.523$   & $-0.182$   &  $-0.147$ &  $6.790$  & T-II\\
     \hline
\end{tabular}
\caption{The different images on the lens plane of points in the source plane}
\label{tab1}
\end{table}

\section{Mapping Magnification}
In this section, the variation of Magnification in the lens plane will be shown. In order to do this the magnification on each point on the lens plane is calculated by taking the reciprocal of $\det(A)$ as suggested in equation \eqref{eq17}. But to show the type of image the modulus of $\det(A)$  is not taken and thus in this graph region where the type I image is formed is represented by red and the region where the type II image is formed is represented by blue. It is clearly visible that this is a somewhat different representation of the critical curve. 

Now the computation of Jacobean was done using the 'FASTELL4py' takes the lens plane coordinates and not the source plane coordinates so it's safe to say in this case the magnification is a function of the lens plane. So a similar plot cannot be done for the source plane. In the last section, it can be noted that the Caustic curve was plotted from the Critical Curve and not directly from the Jacobean. Hence the magnification on the source plane can't be plotted as the Jacobean calculated using the algorithm is only dependent on the coordinates of the lens plane.

\begin{figure}
	\includegraphics[width=\columnwidth]{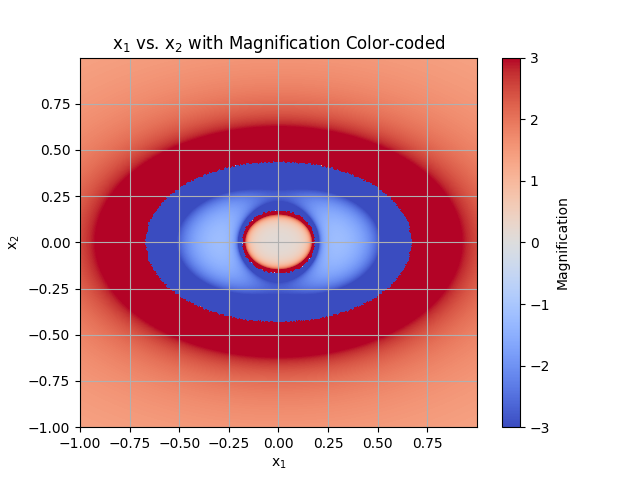}
    \caption{The magnification map on the lens plane using the "coolwarm" style of color mapping. The red shows where there is the positive magnification and the blue where there is negative magnification. The parameters are $q=0.28$, $u=0.020$ and $n=0.69$}
    \label{fig5}
\end{figure}

\section{Comparison of the Critical and the Caustic curves}
\begin{figure}
	\includegraphics[width=\columnwidth]{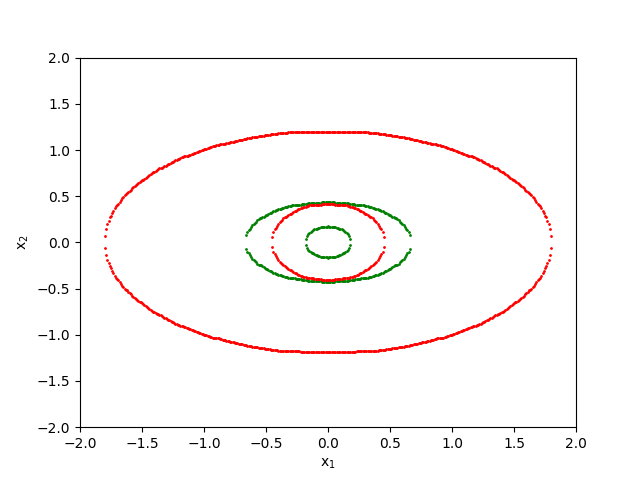}
    \caption{Critical curve for the NFW model is represented in green and that by Bogdanov in red.}
    \label{fig6}
\end{figure}

\begin{figure}
	\includegraphics[width=\columnwidth]{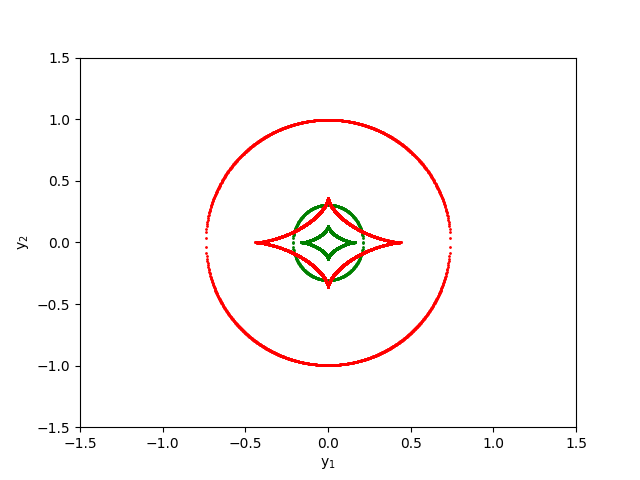}
    \caption{Caustic curve for the NFW model is represented in green and that by Bogdanov in red.}
    \label{fig7}
\end{figure}

The critical curve for the model suggested by Bogdanov \cite{bogdanov2001} is represented in the graph as red and the critical curve for the NFW model is represented in green in fig \ref{fig6}. A similar color scheme is followed in the plot of the caustic curves in fig \ref{fig7}.

The parameters for that of the model suggested by Bogdanov \cite{bogdanov2001} are given in the same paper cited above.  Note that to plot the graphs in this paper the same constants were taken that were used in the paper of Bogdanov \cite{bogdanov2001} i.e. same values of $M_x$, $R_0$ and $R_x$ and using them $\rho_0$ is calculated from equation \eqref{eq3}. Thus, this justifies why the two models can be compared.

\textbf{Analysis from the Graph :}

From figures \ref{fig6} and \ref{fig7} a few things can be analyzed. The most important point to take away is that even though these are very different models they produce critical and caustic curves in the same order even though the one produced by the NFW model occupies a smaller region than that of the one suggested by Bogdanov \cite{bogdanov2001}.

The angular separation in the lens plane in the $x_1$ direction for the model suggested by Bogdanov \cite{bogdanov2001} amounts to $0.00303$ arcsecond while that of the NFW model amounts to about $0.0011$ arcsecond. This clearly shows that they are of the same order. This was calculated in the following formula:

\begin{equation}
    \theta=\frac{R_x}{D_d}x
    \label{eq18}
\end{equation}

Here, $R_x$, $D_d$ has the usual meaning, and $x_1$ is the lens plane coordinate in the x-direction. This calculation clearly shows that both models produce Critical curves of the same order.

Similarly, we can use a similar formula to calculate the angular separation in the source plane. This amounts to $0.0012$ arcsecond for the model suggested by Bogdanov \cite{bogdanov2001} and $0.0003$ arcsecond for the NFW model. To write the formula explicitly:

\begin{equation}
    \beta=\frac{R_x}{D_d}y
    \label{eq19}
\end{equation}
Where all the constants have the usual meaning and $y_1$ is the source plane coordinate in the x-direction. 

\section{Results and Discussion}
The Critical and Caustic curves were plotted along with the deflection angle and how it changes with the x-plane coordinates. Then it was compared to previous model first suggested by Gurevich \cite{gurevich1997} and it was found that the caustic and critical curves covers a much smaller area than their counterpart. This may provide a more accurate representation as NFW model is very accurate in the halo region of the galaxy. This paper deals with lenses in the Galactic Halo of the Milky Way galaxy and the sources in the LMC or the Large Magellanic Cloud. So this paper simulates the small scale dark matter structures that can be present in halo of the Milky Way Galaxy.

Plotting the critical and caustic curves shows the position of points with maximum magnification or position of maximum flux generation and hence is quite an important tool for detection of microlensing events and thus these structure in the Galactic Halo. The angular distance of the two extremities of the caustic as well as the critical curves are quite small and may be challenging to resolve using the telescopes used for such detection. If such angular resolution can be reached then pin pointing it to the exact point in the lens as well as the source plane can be instrumental in detecting events. From the plot of the critical and caustic curves the angular size of the major axis of the caustic and critical and caustic curves were calculated and for the both the cases it was found to be in the milli-arcsecond range. 

The position and movement of the source on the plane of the caustic curves produce a wide array and types of the flux curves which can be picked up by the telescope sensors. Hence the position of the source in the caustic curve is vital. The types of flux curves produced can led to the prediction of several parameters like mass as we have seen in the paper by Calchi \cite{calchi2011microlensing}. These parameters can tell us a lot about our lens for example if it is a MACHO, a binary lens, exoplanet etc.  Using the Caustic and critical curves astronomers can predict the path of the source object(star) across the sky and schedule observations so that it is captured at the right time. This may able to gap one of the major drawbacks of microlensing which is that the background source moves away and hence the even is not just rare but a one time thing. So precisely predicting these curves may prove instrumental in capturing this event at the right time with the best source position possible.

The angular resolution of telescopes dedicated for microlensing expeditions like OGLE which uses 1.3 m Warsaw University Telescope of Las Campanas Observatory in Chile is around a few arcsecond which is not even close to the angular size of the major axis of the caustic and critical curves. This may lead to inaccuracies as it is impossible to resolve the position of the source in reference to the Caustic with such resolution. Modern telescopes with large diameters, these problems can be easily overcome. Modern telescopes with huge sizes have comparable resolution and thus this method of using caustic and critical curves to predict microlensing events can be a fore runner in the discovery of the proposed structures.  

\section{Conclusion}
In this paper the NFW density profile has been applied to small scale clusters of rotating dark matter objects in the sub-solar range. NFW model is a much more preciese dark matter density model than the theoretical model used in the paper \cite{bogdanov2001}. The images produced due to the gravitational lensing cannot be resolved by the telescopes available however these lenses acts as microlenses. The motion of the observer, lens and the source with respect to each other produce a wide number of flux curves. These microlenses produce caustic and critical curves and the movement of the source with respect to the caustic curve produces specific types of flux curves and thus simulation of these caustic and critical curves can make sure that the event is captured in the right time hence better observations. The expeditions that carry on microlensing events like OGLE, MACHO, EROS have telescopes with low resolution as compared to the angular size of the major axis of the caustic as well as critical curves and thus this process may not be as accurate and thus lead to many inaccuracies while taking the observations. 

However with new telescopes being made and also existing telescopes which are large in size have better resolution. Collaboration during important events may play a significant role. This is not so far fetched idea as there had been collaboration between OGLE, Hubble, ALMA and WISE to detect exoplanet orbiting a binary pair of stars \cite{bennett2016}. More and more collaboration and effort in detecting these invisible structure is of great importance for understanding the structure of dark matter in the small as well as large scales. Malbet \cite{malbet2021} in his article poses several questions like “What is the nature of dark matter?” and proposes that microarcsecond relative astrometry in space can answer a large number of open questions. A precise  microarcsecond astrometry can be instrumental in using the model to detect dark matter objects simulated in this paper. 

There are several efforts that is being done for example the ongoing OGLE-IV is still collecting data and is yet to be analyzed for the detection of Dark Matter structures. One of the upcoming missions which will be capable of conducting microlensing efforts is Nancy Grace Roman Space Telescope which is a specialized space telescope for using microlensing. Among other objectives it will look for these small scale clumps of dark matter using microlensing. Hence the application of a more accurate model of dark matter i.e. NFW which produce caustic and critical curves coupled with the more accurate and precise upcoming and modern instruments may prove to be instrumental in detecting small clumps of dark matter and hence give a picture of the small scale local structure of dark matter in our Galaxy.

\section{Acknowledgement}
TRH is grateful to R. Barkana who shared the FASTELL algorithm
and also to Simon Birrer who created a Python wrapper over the Fortran code created by R. Barkana. This work was supported by Amity University, Kolkata and ICARD, Aliah University by providing instrumental encouragement. MK is thankful to the Inter-University Centre for Astronomy and Astrophysics (IUCAA), Pune, India for providing the Visiting Associateship and ICARD, Aliah University for providing the research facilities.

\section{Data Availability Statement}

This manuscript has no associated data or the data will not be deposited. [Authors’ comment: This manuscript has no measured data associated; the plots involve data generated by modelling.]

% Bibliography

%% [A] Recommended: using JHEP.bst file
 \bibliographystyle{JHEP}
 \bibliography{biblio.bib}

%% or
%% [B] Manual formatting (see below)
%% (i) We suggest to always provide author, title and journal data or doi:
%% in short all the informations that clearly identify a document.
%% (ii) please avoid comments such as "For a review'', "For some examples",
%% "and references therein" or move them in the text. In general, please leave only references in the bibliography and move all
%% accessory text in footnotes.
%% (iii) Also, please have only one work for each \bibitem.

\end{document}